\documentclass[a4paper]{article}

\usepackage{INTERSPEECH2020}
\usepackage{amsmath}
\usepackage{amssymb}
\usepackage{float}
\usepackage{listings}
\usepackage{hyperref}

\usepackage[symbol]{footmisc}

\title{Multimodal Inductive Transfer Learning for Detection of Alzheimer's Dementia and its Severity}
\name{Utkarsh Sarawgi\textsuperscript{*}, Wazeer Zulfikar\textsuperscript{*}, Nouran Soliman, Pattie Maes}
\address{
  Massachusetts Institute of Technology}
\email{\{utkarshs, wazeer, nouran, pattie\} @mit.edu}

\begin{document}
\maketitle
\renewcommand{\thefootnote}{\fnsymbol{footnote}}
\footnotetext[1]{Equal Contribution}
\begin{abstract}
Alzheimer's disease is estimated to affect around 50 million people worldwide and is rising rapidly, with a global economic burden of nearly a trillion dollars. This calls for scalable, cost-effective, and robust methods for detection of Alzheimer's dementia (AD). We present a novel architecture that leverages acoustic, cognitive, and linguistic features to form a multimodal ensemble system. It uses specialized artificial neural networks with temporal characteristics to detect AD and its severity, which is reflected through Mini-Mental State Exam (MMSE) scores. We first evaluate it on the ADReSS challenge dataset, which is a subject-independent and balanced dataset matched for age and gender to mitigate biases, and is available through DementiaBank. Our system achieves state-of-the-art test accuracy, precision, recall, and F1-score of 83.3\% each for AD classification, and state-of-the-art test root mean squared error (RMSE) of 4.60 for MMSE score regression. To the best of our knowledge, the system further achieves state-of-the-art AD classification accuracy of 88.0\% when evaluated on the full benchmark DementiaBank Pitt database. Our work highlights the applicability and transferability of spontaneous speech to produce a robust inductive transfer learning model, and demonstrates generalizability through a task-agnostic feature-space. The source code is available at \href{https://github.com/wazeerzulfikar/alzheimers-dementia}{\lstinline|https://github.com/wazeerzulfikar/alzheimers-dementia|}\\
\end{abstract}


\section{Introduction} 
Alzheimer's disease is a progressive disorder that causes brain cells to degenerate and is the most common cause of dementia worldwide. It mainly causes cognitive and behavioural deterioration of the patients \cite{molinuevo2011role} which is reflected through memory loss, language impairment \cite{escobar2010calidad}, and a decreased ability to express their needs. This in turn affects their quality of life, prognosis, and social relationships. Consequently, it has been imposing increased health risks \cite{schulz1999caregiving} and a significant financial burden to patients, caregivers, families, and healthcare institutions \cite{atance2004costs}. The number of people with dementia worldwide in 2015 was estimated at 47.47 million, and reaching 135.46 million in 2050 \cite{prince2013global}. At the time of writing this paper, someone in the U.S. develops Alzheimer’s disease every 66 seconds, and by 2050 it is projected to be 33 seconds \cite{alzheimer20162016}. According to the World Health Organization, the global economic burden is nearly a trillion dollars which amounts to 1.1\% of the global GDP. \cite{worldtop}, with 63\% of people with dementia living in low- and middle-income countries \cite{worldepidemiology}. In this work, we aim to take a significant step towards more reliable, cost-effective, scalable, and noninvasive technologies to detect the onset of Alzheimer's dementia (AD) and predict the Mini-Mental State Exam \cite{tombaugh1992mini} scores to estimate the severity of it.

Dementia can be strongly characterized by cognitive degeneration leading to language impairment which primarily occurs due to decline in semantic and pragmatic levels of language processing \cite{ferris2013language}. It has been widely reported that AD can be more sensitively detected with the help of a linguistic analysis than with other cognitive examinations \cite{szatloczki2015speaking} and also long before the diagnosis is medically confirmed \cite{mesulam2008alzheimer}. The temporal characteristics of spontaneous speech, such as speech tempo, number of pauses in speech, and their length are sensitive detectors of the early stage of the disease \cite{fraser2016linguistic, luz2018method, mirheidari2018detecting, haider2019assessment, pulido2020alzheimer}. Given the relative ease of collecting balanced and representative data of spontaneous speech and their corresponding transcriptions, they can be utilized in early and robust predictions for the onset of AD. 

Consequently, our research work:

\begin{enumerate}
\item
Presents a novel architecture comprising of domain-specific feature engineering and artificial neural networks for Alzheimer's Dementia (AD) detection and its severity through classification and MMSE score regression (Section \ref{mandm}).
\item
Evaluates the system in a subject-independent setting with a carefully curated balanced and stratified dataset matched for age and gender, to help minimize common biases in the tasks (Section \ref{data}).  
\item
Achieves state-of-the-art test accuracy, precision, recall, and F1-score for AD classification, and state-of-the-art test RMSE for MMSE score predictions on the ADReSS (Alzheimer’s Dementia Recognition through Spontaneous Speech) dataset. To the best of our knowledge, the system further achieves state-of-the-art AD classification accuracy when evaluated on the full benchmark DementiaBank Pitt database (Sections \ref{result} and \ref{discuss}).
\item
Spans a multimodal feature space to increase generalizability and robustness, and uses ensemble mechanisms to leverage individual feature sets and model performances.
\item
Reflects upon the transferability and interdependence of the two tasks of AD classification and MMSE regression.
\end{enumerate}

\section{Related work}
Many current AD detection studies use medical imaging \cite{lu2018multimodal, ortiz2018discriminative, sarraf2016deep} with deep neural networks and random forests. Several studies claim that AD can be sensitively detected in early stages by doing linguistic analysis which leverages speech and language features to train machine learning models for the detection of AD \cite{fraser2016linguistic, luz2018method, mirheidari2018detecting, haider2019assessment, pulido2020alzheimer, di2019enriching}. 

In study \cite{rentoumi2014features}, machine learning methods based on image description were used reaching an accuracy of 75\% on a limited number of subjects enrolled in a longitudinal study. Study \cite{liu2020new} used logistic regression trained with spectrogram features extracted from audio files reaching accuracy of 83.3\% and 84.4\% on VBSD and Dem@Care datasets respectively. Data used in each of the above works are limited to around 32 to 36 subjects and highly imbalanced between the classes and across age and gender. In study \cite{luz2018method}, different traditional classification algorithms like logistic regression, SVM, and more were used to learn speech parameters from dialogues in Carolina Conversations Collection. The best of their solutions reached 86.5\% leave-one-out cross-validation (LOOCV) accuracy with 38 subjects. Works based on data extracted from DementiaBank have reported scores of around 0.87, 0.85, 0.82, 0.80, 0.79, 0.64, and 0.62 \cite{kong2019neural, masrani2018detecting, fraser2016linguistic, yancheva2016vector, hernandez2018computer, luz2017longitudinal, luz2020alzheimer} for AD classification. Study \cite{yancheva2015using} used speech related features to get a mean absolute error (MAE) of 3.83 for MMSE scores with longitudinal data derived from DementiaBank. While a number of works have proposed speech and language based approaches to AD recognition through speech, their studies have used different, often unbalanced and acoustically varied data sets, thereby introducing bias and hindering generalization, reproducibility and comparability of the proposed approaches.


\section{Methods and materials} \label{mandm}

\subsection{Dataset} \label{data}
The DementiaBank Pitt database \cite{becker1994natural} consists of speech recordings and transcripts of spoken picture descriptions elicited from participants through the Cookie Theft picture from the Boston Diagnostic Aphasia Exam \cite{goodglass2001bdae}. The database consists of multiple samples per subject corresponding to multiple visits. The full database contains 242 speech samples from 99 control healthy subjects and 255 speech samples from 168 AD subjects. The dataset also provides Mini-Mental Status Examination (MMSE) scores, ranging from 0 to 30, of the subjects, which offers a way to quantify cognitive function and screen for cognitive loss by testing the individuals' orientation, attention, calculation, recall, language and motor skills \cite{tombaugh1992mini}. A 10-fold cross-validation was used on this database for fair comparison with previously reported results.

The ADReSS Challenge Dataset \cite{luz2020alzheimer} is a balanced subset consisting of 156 speech samples, each from a unique subject, matched for age and gender and evenly spread across the two classes, AD and non-AD. A stratified train-test split of around 70-30 (108 and 48 subjects) for this dataset was provided by the challenge. The test set was held out for all experimentation until final evaluation. Any cross-validation mentioned in the paper refers to cross-validation using the train split. Normalized speech segments are also provided, but we only use full audio samples. The MMSE scores provided are used as labels for the regression task.

We first evaluate on the balanced ADReSS dataset and then extend the evaluation to the full DementiaBank Pitt database. 

\subsection{Feature engineering} \label{sec:feature_eng}
People with dementia show symptoms of cognitive decline, impairment in memory, communication, and thinking \cite{pulido2020alzheimer}. To include such domain knowledge and context, our system extracts cognitive and acoustic features using three different strategies, which are then prepared and fed into their respective neural models. Similarly extracted features have been repeatedly used to propose speech recognition based solutions for automated detection of mild cognitive impairment from spontaneous speech \cite{toth2018speech, pulido2020alzheimer}. The following features were extracted upon exploring the data to find the most descriptive set of correlated features for detecting AD and its severity:

\textbullet\hspace{1mm}\textit{Disfluency:} A set of 11 distinct and carefully curated features from the transcripts, like word rate, intervention rate, and different kinds of pause rates reflecting upon speech impediments like slurring and  stuttering. These are normalized by the respective audio lengths and scaled thereafter.

\textbullet\hspace{1mm}\textit{Acoustic:} The ComParE 2013 feature set \cite{eyben2013recent} was extracted from the audio samples using the open-sourced openSMILE v2.1 toolkit, widely used for affect analyses in speech \cite{eyben2010opensmile}. This provides a total of 6,373 features that include energy, MFCC, and voicing related low-level descriptors (LLDs), and other statistical functionals. This feature set encodes changes in speech of a person and has been used as an important noninvasive marker for AD detection \cite{lopez2012alzheimer, luz2020alzheimer}. Our system standardizes this set of features using z-score normalization, and uses principal component analysis (PCA) to project the 6,373 features onto a low-dimensional space of 21 orthogonal features with highest variance. The number of orthogonal features was selected by analyzing the percentage of variance explained by each of the components.

\textbullet\hspace{1mm}\textit{Interventions:} Cognitive features reflect upon potential loss of train of thoughts and context. Our system extracts the sequence of speakers from the transcripts, categorizing it as subject or the interviewer. To accommodate for the variable length of these sequences, they are padded or truncated to length of 32 steps, found upon analyses and tuning of sequence lengths. 

We evaluated each of these features individually and in a combined fashion to highlight the different configurations and compare their performances.

\subsection{Model architecture and training} \label{sec:model_arch}
\begin{figure}[t]
  \centering
  \includegraphics[width=\linewidth]{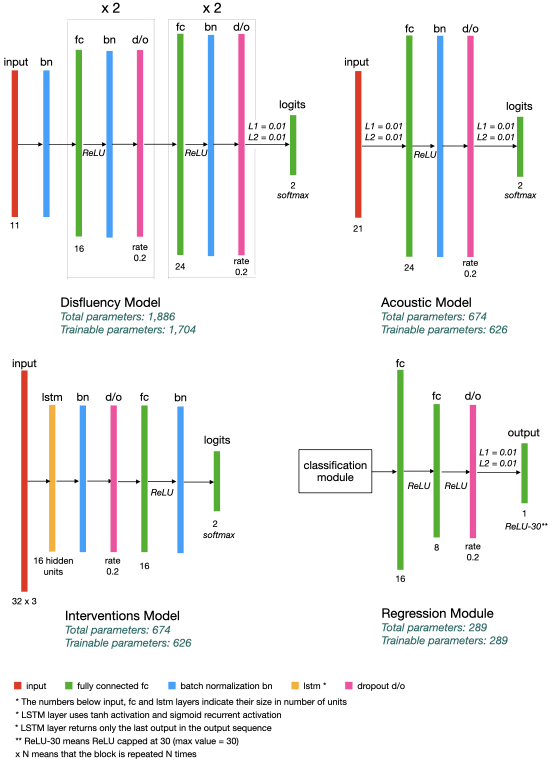}
  \caption{Architecture of (1) Disfluency, (2) Acoustic, (3) Interventions models, and (4) Regression module.}
  \label{fig:model_arch}
  \vspace{-2mm}
\end{figure}

Figure \ref{fig:model_arch} - (1), (2), and (3) illustrate the architecture of the disfluency, acoustic, and interventions models respectively. The disfluency model is a multi-layer perceptron (MLP) that projects the 11-feature input to a higher dimensional space for better separability of the binary classes. The acoustic model is an MLP with a single hidden layer that adds non-linearity and regularizes the PCA decomposed feature space. The interventions model uses a recurrent architecture to learn the temporal relations from the sequence of interventions. These models were trained with corresponding inputs obtained upon feature engineering (Section \ref{sec:feature_eng}), and one-hot encoded binary class labels.

To leverage the features learnt from classification for regression, transfer learning was done on the trained classification models. The regression module, as shown in Figure \ref{fig:model_arch} - (4) replaced the terminal output layer in the models and the remaining original layers were frozen. The resultant models were then trained with MMSE scores as labels.

A 5-fold cross-validation setting was adopted for evaluation. The models were also evaluated in a leave-one-out cross validation (LOOCV) setting, which in the case of ADReSS dataset is equivalent to leave-one-subject-out cross validation (LOSO) since each datapoint is an independent subject. Each training run used a batch size of 8; and Adam optimizer with a learning rate of 0.01 to minimize categorical cross-entropy loss for classification, and a learning rate of 0.001 to minimize mean squared error loss for regression. The best models were saved by monitoring the validation loss in each fold.

To leverage all sets of features and models together, a parallel ensemble was performed using the outputs of the three models for each of the two tasks independently. We experimented with three kinds of ensemble modules for classification:

\textbullet\hspace{1mm}\textit{Hard:} A majority vote was taken between the predictions of the three individual models. 

\textbullet\hspace{1mm}\textit{Soft:} To leverage the confidence of the predictions, a weighted sum of the class probabilities was computed for final decision. The weight used was $1/N$ where $N$ is the total number of models.

\textbullet\hspace{1mm}\textit{Learnt:} Instead of weighing the confidence of all the models equally as in soft voting above, we used a logistic regression to learn the weights. A logistic regression voter was trained using class probabilities as inputs. 

For regression, the predictions of all the individual models were averaged by the ensemble module.

\section{Results} \label{result}
The results of the experiments were recorded using a combination of accuracy, precision, recall and F1-score for classification, and root mean squared error (RMSE) for regression. 

\subsection{ADReSS Challenge dataset}
\begin{table}[h]
    \caption{5-fold cross validation results of the models. Accuracy measures the AD classification performance while RMSE measures the MMSE score regression performance over all 5 folds. Ensemble in this table refers to hard ensemble for classification and the regression ensemble for regression.}
    \label{metrics}
    \centering
    \def\arraystretch{1.2}
    \setlength\tabcolsep{9pt}
    \begin{tabular}{llcc}
    \toprule
      Model & Split & Accuracy & RMSE\\
      \hline
      Disfluency & Train & 0.87 $\pm$ 0.08 & 4.37 $\pm$ 0.40\\
      & Val & 0.89 $\pm$ 0.05 & 4.87 $\pm$ 0.78\\
      \hline
      Acoustic & Train & 0.89 $\pm$ 0.03 & 4.40 $\pm$ 0.64\\
      & Val & 0.83 $\pm$ 0.07 & 5.63 $\pm$ 1.15\\
      \hline
      Interventions & Train & 0.82 $\pm$ 0.06 & 5.05 $\pm$ 0.56\\
      & Val & 0.89 $\pm$ 0.04 & 4.70 $\pm$ 0.96\\
      \hline
      Ensemble & Train & \textbf{0.91 $\pm$ 0.04} & \textbf{3.65 $\pm$ 0.38} \\
      & Val & \textbf{0.92 $\pm$ 0.06} & \textbf{4.26 $\pm$ 0.75}\\
    \bottomrule
    \end{tabular}
\end{table}

\begin{table}[h]
    \caption{5-fold cross-validation accuracies of different ensemble mechanisms for AD classification.}
    \label{ensembletable}
    \centering
    \def\arraystretch{1.1}
    \setlength\tabcolsep{9pt}
    \begin{tabular}{llc}
    \toprule
      Ensemble Type & Split & Accuracy \\
      \hline
      Hard & Train & 0.91 $\pm$ 0.04 \\
      & Val & \textbf{0.92 $\pm$ 0.06} \\
      \hline
      Soft & Train & 0.86 $\pm$ 0.04 \\
      & Val & 0.86 $\pm$ 0.04 \\
      \hline
      Learnt & Train & \textbf{0.95 $\pm$ 0.03} \\
      & Val & 0.81 $\pm$ 0.08\\
    \bottomrule
    \end{tabular}
    \vspace{-3mm}
\end{table}

Table \ref{metrics} shows the 5-fold cross-validation results for the classification task. The individual features achieved competitive performance, although the acoustic model slightly overfits while the interventions model marginally underfits on the data. The ensemble model counteracted these and achieved an increased 5-fold mean training as well as validation accuracy with comparable variance. The low variance generally observed across all runs signifies high model stability across folds which is essential in small datasets. Similar observations can be seen on the regression task in Table \ref{metrics}, where the ensemble model reduced the train and validation mean RMSE as well as the variance. This is consistent with the intuition behind using transfer learning using the trained classification models through the addition of a regression module.

The improvement in performance upon ensembling the three models as compared to the individual models further reflects upon the significance of leveraging acoustic and cognitive features together from multimodal speech and text inputs.

Table \ref{ensembletable} shows the 5-fold cross validation results of different parallel ensemble techniques, discussed in Section \ref{sec:model_arch}, for the classifiation task. The learnt ensemble showed signs of overfitting due to the extra trainable parameters in the model. The soft and hard ensemble helped counter this. However, the hard ensemble proved to be the most competitive by improving training and validation accuracies along with a strong degree of generalization across folds.

\begin{figure}[H]
\vspace{-4mm}
  \centering
  \includegraphics[width=\linewidth]{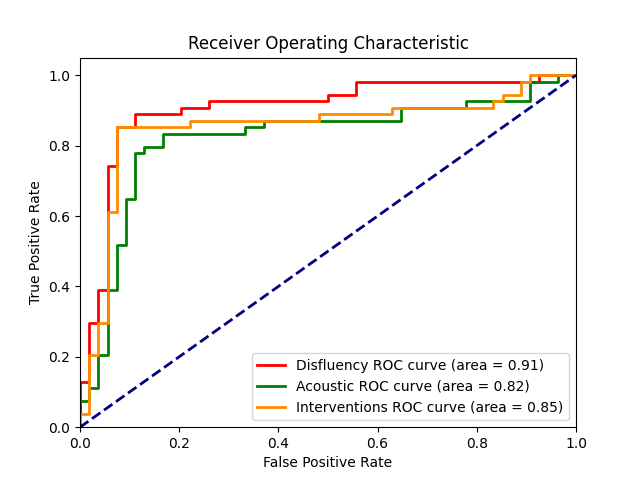}
  \caption{Receiver Operating Characteristic for Disfluency, Acoustic, and Interventions models, cumulatively calculated over validation splits of all the folds of 5-fold cross-validation.}
  \vspace{-2mm}
  \label{fig:roc}
  \vspace{-2mm}
\end{figure}

Figure \ref{fig:roc} shows the receiver operating characteristic (ROC) curve for the individual models on the classification task. The ROC is cumulatively calculated over the validation splits of all 5 folds of cross-validation.

 We compare our results with the currently available baseline performance results on this dataset \cite{luz2020alzheimer}. Amongst our models, the best performing model, the hard ensemble classification model and the ensemble regression model, considerably improved all the metrics on the LOSO as well as the held-out test set on AD classification and regression, as can be seen in Table \ref{baselineclf} and Table \ref{baselinereg} respectively.
 
\begin{figure}[h]
  \centering
  \includegraphics[width=\linewidth]{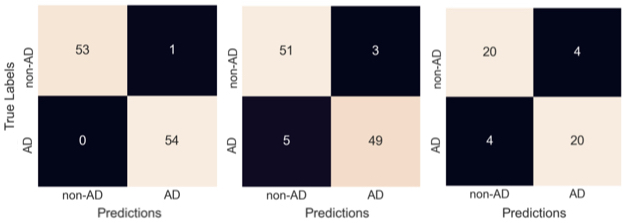}
  \caption{Confusion matrices for the hard ensemble classification model (1) cumulatively calculated over the validation splits of all the folds of LOOCV and (2) 5-fold cross-validation, and (3) calculated on the held out test set.}
  \label{fig:confusion_matrix}
  \vspace{-5mm}
\end{figure}

 The confusion matrices in Figure \ref{fig:confusion_matrix} provide further insights into the predictions of the hard ensemble classification model that has been compared with the baseline in Table \ref{baselineclf}.

\begin{table}[h]
    \caption{Baseline comparison of the AD classification. Our test results below are corresponding to the hard ensemble model.}
    \label{baselineclf}
    \centering
    \def\arraystretch{1.2}
    \setlength\tabcolsep{2pt}
    \begin{tabular}{llccccc}
    \toprule
       & Model & Accuracy & Precision & Recall & F1-Score\\
      \hline
      LOSO & Luz et al. \cite{luz2020alzheimer} & 0.77 & 0.77 & 0.76 & 0.77\\
      &  Ensemble (\textit{ours}) & \textbf{0.99} & \textbf{0.99} & \textbf{1.00} & \textbf{0.99}\\
      \hline
      TEST & Luz et al. \cite{luz2020alzheimer} & 0.75 & 0.83 & 0.62 & 0.71\\
      &  Ensemble (\textit{ours}) & \textbf{0.83} & \textbf{0.83} & \textbf{0.83} & \textbf{0.83}\\
    \bottomrule
    \end{tabular}
    \vspace{-3mm}
\end{table}

\begin{table}[h]
    \caption{Baseline comparison of the MMSE score regression. Our test results are corresponding to the regression ensemble.}
    \label{baselinereg}
    \centering
    \def\arraystretch{1.1}
    \setlength\tabcolsep{11pt}
    \begin{tabular}{llcccc}
    \toprule
       & Model & RMSE\\
      \hline
      LOSO & Luz et al. \cite{luz2020alzheimer} & 4.38\\
      &  Ensemble (\textit{ours}) & \textbf{0.82}\\
      \hline
      TEST & Luz et al. \cite{luz2020alzheimer} & 5.20\\
      &  Ensemble (\textit{ours}) & \textbf{4.60}\\
      
    \bottomrule
    \end{tabular}
    \vspace{-3mm}
\end{table}

\subsection{DementiaBank Pitt database}
\begin{table}[h]
    \caption{Comparison of the AD classification on DementiaBank Pitt. All are 10-fold cross-validation results. Our results below are corresponding to the hard ensemble model.}
    \label{baselinedb}
    \centering
    \def\arraystretch{1.2}
    \setlength\tabcolsep{2pt}
    \begin{tabular}{lccccc}
    \toprule
       Model & Accuracy & Precision & Recall & F1-Score\\
      \hline
      Fraser et al. \cite{fraser2016linguistic} & 0.82 & - & - & -\\
     Masrani \cite{masrani2018detecting} & 0.85 & - & - & 0.85\\
      Kong et al. \cite{kong2019neural} & 0.87 & 0.86 & \textbf{0.91} & \textbf{0.88}\\
      Ensemble (\textit{ours}) & \textbf{0.88} & \textbf{0.92} & 0.82 & \textbf{0.88}\\
    \bottomrule
    \end{tabular}
    \vspace{-3mm}
\end{table}

The same AD classification models were retrained on the DementiaBank Pitt database and a 10-fold cross-validation was performed for fair comparison with previously reported results. To the best of our knowledge, our hard ensemble model achieves state-of-the-art 0.88 $\pm$ 0.04 accuracy, also showing minimal variance across the folds (Table \ref{baselinedb}). 

\section{Discussion and Future Work} \label{discuss}

There has been substantial work using spontaneous speech samples and manual transcriptions present in the DementiaBank dataset \cite{becker1994natural}. Some of the highest reported scores for AD classification are 0.87, 0.85, 0.82, 0.80, 0.79, 0.64, and 0.63 \cite{kong2019neural, masrani2018detecting, fraser2016linguistic, yancheva2016vector, hernandez2018computer, luz2017longitudinal, luz2020alzheimer}. Many of these previous results were obtained on datasets with variable subject dependencies. In such datasets, a data point corresponds to a session and there can exist multiple sessions per subject. Given the subject independent setting in ADReSS dataset, our LOSO method clearly distinguishes the left-out test subject. Hence, the near perfect LOSO results on classification and regression (Tables \ref{baselineclf} and \ref{baselinereg}) demonstrate that every subject individually can be correctly evaluated with the engineered features. Furthermore, almost all previous results are reported using cross-validation, whereas our work is evaluated on a designated held-out test set as well. This helps overcome `validation overfitting' which is prone in small dataset settings. 

Study \cite{yancheva2015using} used speech related features to obtain a cross-validated mean absolute error (MAE) of 3.83 for MMSE scores with data derived from DementiaBank. Our ensemble regression model recorded a cross-validated MAE of 3.01 on ADReSS dataset.

Through considerable improvements in both the AD classification and MMSE score regression by employing an ensemble of independent models extracting acoustic and cognitive features, our work reveals the potential of multimodal analysis and its applicability to a age and gender balanced subject-independent dataset. Future work would include incorporating automated transcription of speech samples in our system. The continuous range of the MMSE scores can provide more insights into progression of dementia. This can further be leveraged for risk stratification and analyzing potential causal relationships modelling AD with its symptoms and markers, through a longitudinal dataset.

\section{Conclusion}

We present a novel architecture that uses domain knowledge for inductive transfer learning for AD classification and MMSE score regression. Our work achieves state-of-the-art accuracy, precision, recall, and F1-score of 83.3\% each for AD classification, and state-of-the-art RMSE of 4.60 for MMSE predictions on the designated held-out test set of the ADReSS challenge. To the best of our knowledge, the system further achieves state-of-the-art AD classification accuracy of 88.0\% when evaluated on the full benchmark DementiaBank Pitt database. Our system spans a multimodal feature space to increase generalization and robustness. We aim to extend our work by adding automated transcription, further textual analysis, and personalized context through longitudinal data.  



\bibliographystyle{IEEEtran}
\bibliography{mybib}
\end{document}